\documentclass[sigconf]{acmart}

\usepackage{multirow}
\usepackage{todonotes}
\usepackage{enumitem}
\usepackage{hyperref}
\usepackage{subcaption}
\usepackage{amsmath} 
\usepackage{booktabs}
\usepackage{graphicx}
\usepackage{makecell}  
\usepackage{tabularx} 
\newcolumntype{Y}{>{\centering\arraybackslash}X}

\copyrightyear{2025}
\acmYear{2025}
\setcopyright{cc}
\setcctype{by}
\acmConference[ICAIF '25]{6th ACM International Conference on AI in Finance}{November 15--18, 2025}{Singapore, Singapore}
\acmBooktitle{6th ACM International Conference on AI in Finance (ICAIF '25), November 15--18, 2025, Singapore, Singapore}\acmDOI{10.1145/3768292.3770388}
\acmISBN{979-8-4007-2220-2/2025/11}

\AtBeginDocument{%
  }

\begin{document}


\title{Multi-Agent Reinforcement Learning for Market Making: Competition without Collusion}

%% "author" 

\author{Ziyi Wang}
\email{ziyi.1.wang@kcl.ac.uk}
\affiliation{%
  \institution{King's College London}
  \city{London}
  \country{United Kingdom}
}

\author{Carmine Ventre}
\email{carmine.ventre@kcl.ac.uk}
\affiliation{%
  \institution{King's College London}
  \city{London}
  \country{United Kingdom}
\email{carmine.ventre@kcl.ac.uk}
}

\author{Maria Polukarov}
\email{maria.polukarov@kcl.ac.uk}
\affiliation{%
  \institution{King's College London}
  \city{London}
  \country{United Kingdom}
\email{maria.polukarov@kcl.ac.uk}
}

\renewcommand{\shortauthors}{Wang et al.}

\begin{abstract}
Algorithmic collusion has emerged as a central question in AI: Will the interaction between different AI agents deployed in markets lead to collusion? More generally, understanding how emergent behavior, be it a cartel or market dominance from more advanced bots, affects the market overall is an important research question. 

We propose a hierarchical multi-agent reinforcement learning framework to study algorithmic collusion in market making. The framework includes a self-interested market maker (Agent~A), which is trained in an uncertain environment shaped by an adversary, and three bottom-layer competitors: the self-interested Agent~B1 (whose objective is to maximize its own PnL), the competitive Agent~B2 (whose objective is to minimize the PnL of its opponent), and the hybrid Agent~B$^\star$, which can modulate between the behavior of the other two. To analyze how these agents shape the behavior of each other and affect market outcomes, we propose interaction-level metrics that quantify behavioral asymmetry and system-level dynamics, while providing signals potentially indicative of emergent interaction patterns.

Experimental results show that Agent~B2 secures dominant performance in a zero-sum setting against B1, aggressively capturing order flow while tightening average spreads, thus improving market execution efficiency. In contrast, Agent~B$^\star$ exhibits a self-interested inclination when co-existing with other profit-seeking agents, securing dominant market share through adaptive quoting, yet exerting a milder adverse impact on the rewards of Agents~A and B1 compared to B2. These findings suggest that adaptive incentive control supports more sustainable strategic co-existence in heterogeneous agent environments and offers a structured lens for evaluating behavioral design in algorithmic trading systems.
\end{abstract}

%% http://dl.acm.org/ccs.cfm.

%% Keywords.

%%\received{20 February 2007}
%%\received[revised]{12 March 2009}
%%\received[accepted]{5 June 2009}

\maketitle

\section{Introduction}

Modern electronic markets are composed of highly heterogeneous trading agents operating under uncertainty, informational asymmetry, and execution competition. The interactions among these agents not only shape price formation and liquidity, but also determine broader dynamics of stability and market structure. Understanding how strategic heterogeneity influences these outcomes is crucial for modeling intelligent financial systems.

Multi-agent reinforcement learning (MARL) has gained increasing attention in market modeling, offering tools for simulating strategic interactions under uncertainty. However, how heterogeneous agents adapt to diverse counterparts and shape market outcomes remains insufficiently understood.

\begin{figure}[htbp]
  \centering
  \includegraphics[width=\linewidth]{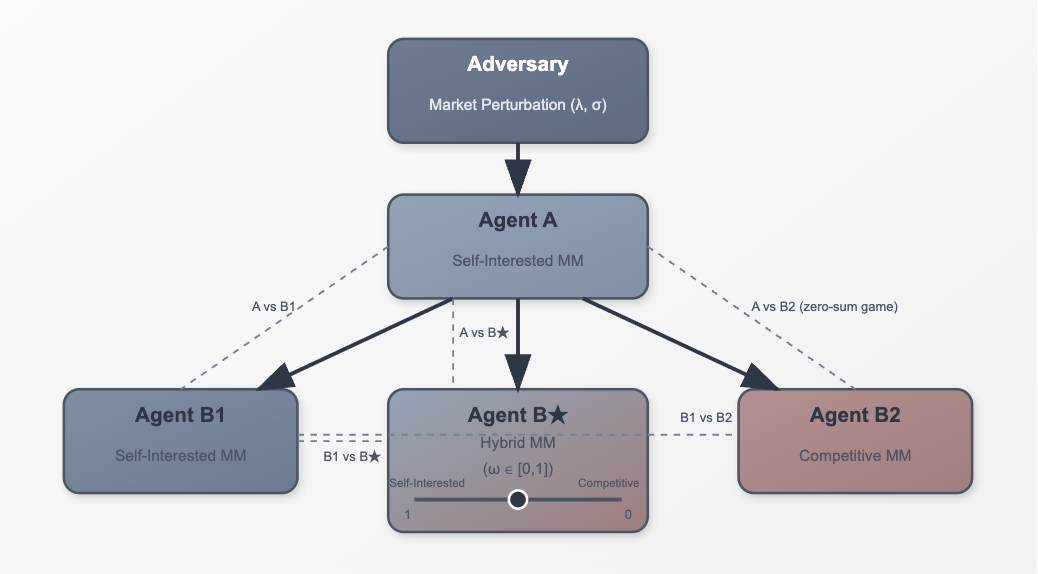}
  \caption{Hierarchical multi-agent architecture with a hybrid agent modulating between self-interested and competitive roles.}
  \label{fig:architecture}
\end{figure}

To address this, we propose a \textit{hierarchical, behavior-structured multi-agent learning framework} for modeling agent-level heterogeneity and its impact on market interactions. The system comprises three functional layers: (i) a top-layer adversary, trained via reinforcement learning, perturbs volatility and order arrival processes to simulate external uncertainty; (ii) a mid-layer Agent~A, representing a self-interested market maker trained to maximize profit under adversarial stress; and (iii) a bottom layer of competitors—Agent~B1 is a self-interested market maker optimizing for individual profit; Agent~B2 is a competitive agent trained in a zero-sum setting to minimize Agent~A’s reward; and the hybrid agent B$^\star$ introduces a learnable modulation parameter $\omega \in [0,1]$, enabling dynamic interpolation between self-interested and competitive behaviors. This setup supports controlled analysis of behavioral differentiation across self-interested, competitive, and hybrid agents. The overall agent architecture and strategic interactions are illustrated in Figure~\ref{fig:architecture}.

To measure agent interaction patterns, we introduce a suite of structural metrics capturing synchronous quoting, price similarity, inventory divergence, joint drawdowns, and fill overlaps. These interpretable signals enable the analysis of behavioral convergence, role differentiation, and potential coordination in competitive environments.

Empirical results show that Agent~B2, a competitive market maker facing the self-interested B1 in a zero-sum setting, achieves strong profit through rigid suppression, but at the cost of increased market share concentration and strategic crowding, potentially limiting diversity in liquidity provision. In contrast, Agent~B$^\star$—despite exhibiting a self-interested inclination when facing multiple profit-seeking agents—demonstrates adaptive flexibility. It successfully captures greater market share while maintaining stable returns and imposing milder pressure on overall market dynamics. In heterogeneous agent markets, such adaptive behavior suggests a more sustainable and strategically robust pathway for long-term coexistence.

\section{Related Work}
Agent-based modeling (ABM) has long served as a foundation for simulating financial systems, offering a principled way to capture decentralized interaction, adaptive decision-making, and systemic risk propagation~\cite{lebaron2006agent, lux1999scaling, foucault2013market}. Recent applications of ABM highlight its value as a computational approach grounded in the institutional and behavioral structure of financial systems, enabling controlled experimentation on market dynamics and agent heterogeneity~\cite{axtell2025abm, boe2025abm}. This modeling paradigm supports the explicit specification of heterogeneous agent roles, making it well suited to studying emergent market properties that arise from local agent interactions—such as volatility clustering or liquidity fragmentation—rather than from global coordination.

Alongside these developments, advances in reinforcement learning (RL) have enabled dynamic adaptation in market-making agents. In particular, multi-agent reinforcement learning (MARL) extends classical RL to settings of strategic interdependence, facilitating the simulation of execution competition and behavioral feedback~\cite{ganesh2019reinforcement, patel2018optimizing}. To manage complexity, hierarchical RL (HRL) structures decision-making across abstraction levels~\cite{pateria2021hierarchical_survey}, and adversarial RL (ARL) improves robustness by training against perturbed dynamics or hostile opponents~\cite{pinto2017robust_adversarial_rl, spooner2022robust,wang2023robust, wang2024arl}. Our framework builds on these principles by embedding learning agents within a layered architecture, where a non-trading adversary reshapes the market environment, and competing market makers face heterogeneous and evolving strategic pressures.

Understanding the systemic effects of local adaptation remains an open challenge in MARL. Prior studies have noted that emergent market phenomena—such as herding, synchronization, or predatory trading—can amplify fragility even without explicit coordination~\cite{farmer2012complex, axtell2025abm}. To monitor these effects, researchers have explored structural risk indicators such as joint drawdowns, behavioral clustering, or correlated inventory shifts~\cite{battiston2016complexity, bookstaber2017end, bisias2012survey}. Our work contributes to this effort by proposing interpretable, interaction-level metrics for evaluating heterogeneous agent dynamics under asymmetric information and adversarial stress. This supports a behavioral lens on systemic risk, moving beyond aggregate PnL toward a more structural understanding of multi-agent influence in algorithmic markets.

\section{Multi-Layered Agent Design}
\subsection{Market Environment}

We construct a synthetic high-frequency trading environment to simulate interactions between market-making agents and a dynamic order flow. The environment captures key microstructural elements, including price evolution, stochastic order flow, probabilistic execution, and inventory-constrained risk management.

\textbf{Asset Price Dynamics.} The mid-price \( P_t \in \mathbb{R} \) evolves via a discrete-time Brownian motion with drift:
\[
P_{t+1} = P_t + \mu \Delta t + \sigma \epsilon_t, \quad \epsilon_t \sim \mathcal{N}(0, \Delta t)
\]
where \( \mu \) is the drift and \( \sigma \) is the volatility. This formulation is consistent with classical models for electronic market microstructure~\cite{avellaneda2008high}.

\textbf{Order Flow and Matching.} At each step, market orders arrive according to a Poisson process with rate \( \lambda \). Each order batch of size \( N_t \sim \text{Poisson}(\lambda \cdot \Delta t) \) is deterministically split into buy and sell orders: \( B_t = \lfloor N_t / 2 \rfloor \), \( S_t = N_t - B_t \). This ensures minimal directional imbalance while reflecting volume-based anonymous flow.

Market-making agents submit bid and ask quotes as price offsets:
\[
q^{\text{bid}} = P_t - d_t^{\text{bid}}, \quad q^{\text{ask}} = P_t + d_t^{\text{ask}}, \quad d_t^{\text{bid}}, d_t^{\text{ask}} \in [0, d_{\max}].
\]

\textbf{Fill Probability and Execution Allocation.}
Market order execution is modeled as a probabilistic process governed by quote aggressiveness. For a quote placed at offset \( d \) from the mid-price \( P_t \), the fill probability is given by:
\[
p_{\text{fill}}(d) =
\begin{cases}
\exp(-\alpha d), & 0 \le d \le d_{\max} \\
0, & \text{otherwise}
\end{cases}
\]

where \( \alpha > 0 \) is a decay coefficient and \( d_{\max} \) sets a maximum execution range. This reflects empirical findings that more competitive quotes—closer to the mid-price—receive higher execution priority~\cite{cartea2015algorithmic}.

Given \( B_t \) buy and \( S_t \) sell market orders, fills are allocated to agents via a multinomial mechanism weighted by their normalized fill probabilities. For ask-side execution:
\[
\text{Fills}_{\text{ask}} \sim \text{Multinomial}\left(B_t, \left\{ \frac{p_i^{\text{ask}}}{\sum_j p_j^{\text{ask}}} \right\}_i \right)
\]

\noindent
An analogous procedure applies to bid-side fills. Each agent imposes per-side inventory limits:
\[
\text{Fill}_{i}^{\text{bid}} \leq \text{Max}_{\text{bid}}, \quad \text{Fill}_{i}^{\text{ask}} \leq \text{Max}_{\text{ask}}
\]

These constraints prevent excessive position accumulation and enforce realistic market risk bounds.

\textbf{Position Update and PnL.} An agent maintains inventory \( I_t \) and cash \( C_t \), which are updated upon execution as:
\[
\begin{aligned}
I_{t+1} &= I_t + \text{Fills}^{\text{bid}} - \text{Fills}^{\text{ask}} \\
C_{t+1} &= C_t - \text{Fills}^{\text{bid}} \cdot q^{\text{bid}} + \text{Fills}^{\text{ask}} \cdot q^{\text{ask}}
\end{aligned}
\]

The mark-to-market profit is:
\[
\text{PnL}_t = C_t + I_t \cdot P_t
\]

This environment enables strategic decision-making by exposing agents to execution competition, price uncertainty, and inventory risk under realistic market frictions.

\subsection{Agent Architecture and Roles}

We adopt a hierarchical multi-agent framework composed of three functional layers, each designed to capture a distinct strategic role in the market simulation. This layered architecture enables the modeling of exogenous shocks, adaptive policy learning, and emergent agent-agent interactions under shared market dynamics.

\textbf{Adversary (Top Layer).} The top-layer agent does not trade but perturbs the environment by adjusting order flow intensity (\( \lambda \)) and mid-price volatility (\( \sigma \)). It learns to generate adverse conditions through environment-level actions, representing external market stress, regulatory interventions, or structural uncertainty.

\textbf{Agent A (Mid Layer).} Positioned as the core learning market maker, Agent A interacts directly with the adversary-shaped environment. Its goal is to maintain profitability and stability across volatile regimes. Trained in isolation before interacting with other agents, it serves as a robust policy baseline for strategic evaluation.

\textbf{Agents B1, B2, B$^\star$ (Bottom Layer).} The bottom layer comprises agents that interact with a fixed Agent A during multi-agent evaluation. B1 is a self-interested market maker trained in an environment where Agent A follows a pre-trained policy. Its objective is to optimize its own profitability under execution competition. B2 represents a zero-sum adversary whose objective is to suppress A’s performance. B$^\star$ generalizes between these extremes by dynamically modulating its behavior along the self-interested versus opponent-suppression axis. This setup supports analysis of competition, coordination, and behavioral adaptation in heterogeneous agent populations, extending prior work in strategic multi-agent markets.

\subsection{Agent Objectives and Incentive Structure}
Each agent interacts with the environment through a structured observation-action interface, summarized in Table~\ref{tab:agent_obs_action_space}. Observation vectors encode local private state (inventory and cash) as well as global market information (price, time). Agents do not observe any information about other agents’ states or actions, ensuring an asymmetric information environment consistent with decentralized market assumptions~\cite{cartea2015algorithmic}. Actions represent either quoting behavior or environmental control, depending on the agent type.

\begin{table}[htbp]
\small
\centering
\caption{Observation and action space of agents}
\label{tab:agent_obs_action_space}
\begin{tabularx}{\linewidth}{Y Y Y}
\toprule
\textbf{Agent} & \textbf{Observation} & \textbf{Action} \\
\midrule
Adversary & \( (P_t, t) \) & \( (\lambda_t, \sigma_t) \) \\
Agent A & \( (P_t, t, I_t^A, C_t^A) \) & \( (d_t^{\text{bid}}, d_t^{\text{ask}}) \) \\
Agent B1 & \( (P_t, t, I_t^{B1}, C_t^{B1}) \) & \( (d_t^{\text{bid}}, d_t^{\text{ask}}) \) \\
Agent B2 & \( (P_t, t, I_t^{B2}, C_t^{B2}) \) & \( (d_t^{\text{bid}}, d_t^{\text{ask}}) \) \\
Agent B$^\star$ & \( (P_t, t, I_t^{B^\star}, C_t^{B^\star}) \) & \( (d_t^{\text{bid}}, d_t^{\text{ask}}, \omega_t) \) \\
\bottomrule
\end{tabularx}
\end{table}

\textbf{Adversary.}
The adversary operates at the environment level, selecting the order flow intensity $\lambda_t$ and the mid-price volatility $\sigma_t$ within prescribed bounds (cf.\ Section~3.1).
The environment also includes a benchmark market-making module that generates quotes according to a fixed, non-learned linear-utility rule; quoting is thus part of the environment dynamics and lies outside the adversary’s action space.
The adversary’s objective is defined as the negative of the inventory-regularized profit used above, evaluated on the benchmark module:
\[
r_t^{\text{adv}}
= -\Big(\Delta \text{PnL}_t \;-\; \zeta\, I_t^{2} \;-\; \eta\,\mathbb{I}_{\text{terminal}}\, I_t^{2}\Big),
\]
where $\Delta \text{PnL}_t$ and $I_t$ refer to the benchmark module’s mark-to-market change and inventory at time $t$.
Maximizing $r_t^{\text{adv}}$ therefore encourages selections of $(\lambda_t,\sigma_t)$ that degrade the benchmark’s performance, assigning the adversary a zero-sum role relative to profit-seeking agents.

\textbf{Agent A and B1.} These agents receive rewards based on risk-adjusted profitability:
\[
r_t = \Delta \text{PnL}_t - \zeta I_t^2 - \eta \cdot \mathbb{I}_{\text{terminal}} \cdot I_t^2
\]
where \( \Delta \text{PnL}_t \) denotes mark-to-market gain at time \( t \), and \( \zeta, \eta \) are inventory regularization coefficients. This design encourages execution while discouraging large directional positions~\cite{avellaneda2008high, spooner2022robust}.

\textbf{Agent B2.} Operates under a zero-sum formulation. Its objective is to suppress Agent A's performance via inverted reward:
\[
r_t^{\text{B2}} = -r_t^{\text{A}}.
\]

\textbf{Hybrid Agent B$^\star$.} Balances between self-interested and competitive objectives via a learnable modulation parameter $\omega_t \in [0, 1]$. Its reward function is:
\[
r_t^{B^\star} = \omega_t \cdot r_t^{\text{self}} - (1 - \omega_t) \cdot r_t^{A} - \texttt{penalty\_coeff} \cdot (\omega_t - 0.5)^2.
\]
Here, \( r_t^{\text{self}} \) follows the same inventory-penalized objective used by Agent A. The first two terms define a tunable trade-off between self-interested and suppression of A. The final penalty term discourages extreme values of \( \omega_t \), which would otherwise cause the agent to collapse into purely self-interested (\( \omega_t \to 1 \)) or opponent-suppression (\( \omega_t \to 0 \)) regimes. Without this regularization, early-stage learning may overfit to local reward gradients, reducing strategic flexibility and generalization. The penalty anchors B$^\star$ near a neutral stance and enables deviation only when sufficiently justified by the environment.

\subsection{Three-Stage Training Process}
All learning-based agents are trained using proximal policy optimization (PPO), implemented via the Stable-Baselines3 library. Agents A, B1, and B2 follow standard actor-critic architectures, while Agent B$^\star$ employs a custom policy head to support its extended action space and strategic modulation.

The policy of B$^\star$ outputs \((d^{\text{bid}}, d^{\text{ask}}, \omega)\), a three-dimensional action tuple where \( \omega \in [0,1] \), controls the trade-off between self-interested and suppression. A sigmoid activation ensures boundedness, and the output bias on \( \omega \) is initialized to zero, yielding a neutral starting point (\( \omega \approx 0.5 \)). To promote training stability, raw actions are clamped within a bounded range, and the hybrid agent’s modulation parameter is regularized to prevent early collapse into extreme strategies.

Training is organized into three sequential stages: (i) pretraining the adversary to perturb the environment, (ii) training Agent A as a reinforcement learning-based market maker under adversarial conditions, and (iii) independently training B-type agents against a pre-trained Agent A strategy.

\section{Evaluation Metrics}

To assess the effectiveness and impact of learned policies, we adopt a set of performance metrics spanning agent behavior, market outcomes, and systemic risk. These metrics provide a quantitative basis for comparing agents such as B1, B2, and B$^\star$ in heterogeneous competitive settings.

%\subsection
\noindent \textbf{Agent-Level Performance Metrics.} This group captures individual agent behavior, including profitability, risk exposure, market participation, and quoting style. The metrics listed in Table~\ref{tab:agent_metrics} are computed per agent and aggregated across evaluation episodes.
\begin{table}[htbp]
\scriptsize
\centering
\caption{Agent-Level Performance Metrics}
\label{tab:agent_metrics}
\begin{tabularx}{\linewidth}{l X}
\toprule
\textbf{Metric} & \textbf{Mathematical Definition} \\
\midrule
Mean PnL & \( \text{PnL}_{\text{mean}} = \mathbb{E}[\text{PnL}_t] \) \\
PnL Std & \( \text{PnL}_{\text{std}} = \sqrt{\mathbb{V}[\text{PnL}_t]} \) \\
Sharpe Ratio & \( \text{Sharpe Ratio} = \frac{\mathbb{E}[\text{PnL}_t]}{\sqrt{\mathbb{V}[\text{PnL}_t]} + 10^{-6}} \) \\
Inventory Volatility & \( \text{Inv}_{\text{std}} = \sqrt{\mathbb{V}[I_t]} \) \\
Quote Aggressiveness & \( \text{Agg} = \mathbb{E}[(d^{\text{bid}}_t + d^{\text{ask}}_t)/2] \) \\
Market Share & \( \text{Share} = \frac{\text{Total Fills}}{\text{Total Arrivals} + 10^{-6}} \) \\
\bottomrule
\end{tabularx}
\end{table}

%\subsection
\noindent \textbf{Market-Level Outcomes.} These metrics characterize how the quoting behavior of agents shapes the market environment. They are designed to reflect overall market quality, including pricing efficiency, liquidity access, and trading activity. While the formulas in Table~\ref{tab:market_metrics} use Agents A and B for illustration, the definitions are applicable to any pair of market participants.
\begin{table}[htbp]
\scriptsize
\centering
\caption{Market-Level Metrics}
\label{tab:market_metrics}
\begin{tabularx}{\linewidth}{>{\raggedright\arraybackslash}p{0.28\linewidth} >{\raggedright\arraybackslash}X}
\toprule
\textbf{Metric} & \textbf{Mathematical Definition} \\
\midrule
Average Spread & \( \frac{1}{T} \sum_{t=1}^T \left( q^{\text{ask},A}_t - q^{\text{bid},A}_t + q^{\text{ask},B}_t - q^{\text{bid},B}_t \right) / 2 \) \\
Fill Ratio & \( \text{Fill Ratio}^{(i)} = \frac{\sum_t \text{Fills}^{(i)}_t}{\sum_t \text{Arrivals}_t + 10^{-6}} \), for \( i \in \{A, B\} \) \\
Zero-Fill Steps & \( \text{Zero-Fill Count}^{(i)} = \sum_t \mathbb{I}(\text{Fills}^{(i)}_t = 0) \), for \( i \in \{A, B\} \) \\
\bottomrule
\end{tabularx}
\end{table}

%\subsection
\noindent \textbf{Multi-Agent Interaction Metrics.} This class of metrics captures structural interaction patterns among learning agents. They quantify synchronous quoting activity, quote similarity, inventory divergence, joint drawdowns, and fill overlaps. These indicators support the analysis of market segmentation, behavioral convergence, and agent specialization in multi-agent settings.
\begin{table}[htbp]
\scriptsize
\centering
\caption{Multi-Agent Interaction Metrics}
\label{tab:systemic_metrics}
\begin{tabularx}{\linewidth}{>{\raggedright\arraybackslash}p{0.28\linewidth} >{\raggedright\arraybackslash}X}
\toprule
\textbf{Metric} & \textbf{Mathematical Definition} \\
\midrule
Joint Drawdown Ratio & \( \frac{1}{T} \sum_t \mathbb{I}(r_t^{A} < 0 \land r_t^{B} < 0) \) \\
Herding Ratio & \( \frac{1}{T} \sum_t \mathbb{I}\left(|q_t^{\text{bid},A} - q_t^{\text{bid},B}| < \varepsilon \land |q_t^{\text{ask},A} - q_t^{\text{ask},B}| < \varepsilon\right) \) \\
Inventory Divergence & \( \sqrt{\mathbb{V}[I_t^{A} - I_t^{B}]} \) \\
Quote Distance (Bid) & \( \mathbb{E}[|q_t^{\text{bid},A} - q_t^{\text{bid},B}|] \) \\
Quote Distance (Ask) & \( \mathbb{E}[|q_t^{\text{ask},A} - q_t^{\text{ask},B}|] \) \\
Fill Overlap Ratio & \( \frac{1}{T} \sum_t \mathbb{I}[(\text{Fills}^{\text{bid},A}_t > 0 \land \text{Fills}^{\text{bid},B}_t > 0) \lor (\text{Fills}^{\text{ask},A}_t > 0 \land \text{Fills}^{\text{ask},B}_t > 0)] \) \\
\bottomrule
\end{tabularx}
\end{table}

The definitions in Table~\ref{tab:systemic_metrics} use Agents A and B for illustration but are applicable to any agent pair.

\section{Strategic Evaluation and Behavioral Analysis in Multi-Agent Markets}

\subsection{Single-Agent Evaluation of Robustness and Generalization}
In this section, we aim to evaluate the robustness and generalization capacity of learned market-making agents under different training and evaluation environments. Specifically, we examine whether Agent A—adversarially trained through direct interaction with an adversary—can maintain strong performance when deployed in a fixed market (\(\lambda=400\), \(\sigma=1.1\)). We also assess whether Agent B1, which was trained by interacting with a well-trained Agent A but never exposed to adversarial conditions during training, can adapt when evaluated in an adversarially-driven market. Furthermore, we compare the effects of \textit{direct} versus \textit{indirect} exposure to adversarial environments on strategic robustness.

\begin{table}[htbp]
\centering
\caption{Single-Agent Evaluation for Agent A and B1 under Fixed vs. Adversarial Conditions}
\label{tab:single_agent_eval}
\resizebox{\linewidth}{!}{%
\begin{tabular}{c c c c c}
\toprule
\textbf{Metric} & \textbf{A\_Fix} & \textbf{A\_Adversarial} & \textbf{B1\_Fix} & \textbf{B1\_Adversarial} \\
\midrule
PnL\_mean              & 506.490 ± 35.698  & 626.421 ± 40.250  & 437.293 ± 32.030  & 538.535 ± 37.232  \\
Sharpe\_ratio          & 1.749 ± 0.089     & 1.746 ± 0.080     & 1.742 ± 0.093     & 1.741 ± 0.086     \\
Inventory\_volatility  & 28.335 ± 2.383    & 28.142 ± 2.443    & 28.335 ± 2.383    & 28.143 ± 2.443    \\
Quote\_aggressiveness  & 2.480 ± 0.002     & 2.480 ± 0.003     & 2.095 ± 0.004     & 2.095 ± 0.005     \\
Market\_share          & 0.997 ± 0.004     & 0.992 ± 0.005     & 0.995 ± 0.004     & 0.990 ± 0.005     \\
Avg\_spread            & 4.959 ± 0.004     & 4.960 ± 0.005     & 4.189 ± 0.009     & 4.190 ± 0.010     \\
Price\_volatility      & 0.402 ± 0.155     & 0.531 ± 0.205     & 0.402 ± 0.155     & 0.531 ± 0.205     \\
Zero\_fill\_steps      & 27.000 ± 4.273    & 16.870 ± 3.681    & 27.000 ± 4.273    & 16.870 ± 3.681    \\
\bottomrule
\end{tabular}
}
\end{table}

Table~\ref{tab:single_agent_eval} summarizes the results of single-agent evaluations for two representative agents: Agent A and Agent B1. Each agent is tested in two market regimes. In the \textit{Fix} setting, the agent runs alone in a static market with parameters fixed at \(\lambda = 400\) and \(\sigma = 1.1\). In the \textit{Adversarial} setting, the agent also runs alone, but the market parameters are dynamically controlled by a trained adversary, with \(\lambda\) varying within \([300, 500]\) and \(\sigma\) within \([0.2, 2.0]\).

\subsubsection{Comparison 1: Agent A – Fixed vs Adversarial Market}

Agent A achieves a notable increase of approximately 24\% in mean PnL under adversarial conditions, indicating that the policy is capable of exploiting profit opportunities in dynamically changing environments. The Sharpe ratio remains stable, and inventory volatility shows no significant change, suggesting consistently strong risk control. Even as price volatility increases due to adversarial interventions, quote aggressiveness and average spread remain stable across both settings, indicating consistent quoting behavior. Zero-fill steps are substantially lower under adversarial conditions, suggesting either improved agent responsiveness or more abundant trading opportunities due to higher order flow.

Overall, these results suggest that Agent A's strategy demonstrates strong generalization and maintains consistent performance across both fixed and adversarial environments.

\subsubsection{Comparison 2: Direct vs Indirect Exposure to Adversarial Conditions (B1\_Fix vs. B1\_Adversarial \& A\_Adversarial vs. B1\_Adversarial)}

We now analyze whether indirect exposure to adversarial environments through co-training with a robust agent can induce adaptive behavior, and how it compares to direct adversarial training.

Comparing the \texttt{B1\_Fix} and \texttt{B1\_Adversarial} columns in Table~\ref{tab:single_agent_eval}, we observe that Agent B1, despite never being exposed to adversarial conditions during training, achieves a significantly higher PnL when evaluated under adversarial market dynamics. Execution efficiency improves substantially, with zero-fill steps dropping from 27 to 16.87, even though Sharpe ratio, inventory volatility, and quote aggressiveness remain stable. This suggests that B1 maintains a consistent quoting style and risk posture, yet still achieves better execution under volatile market dynamics. The enhanced fill rate, without changes in quoting parameters, implies that B1 has learned to position its quotes more effectively in response to fluctuating order flow—likely by indirectly adapting through repeated interaction with the robust Agent A during training. This suggests the emergence of indirect robustness propagation through structural co-learning.

Comparing the \texttt{A\_Adversarial} and \texttt{B1\_Adversarial} columns in Table~\ref{tab:single_agent_eval}, we find that Agent~A achieves a significantly higher PnL (approximately 16\% above B1) despite quoting with larger average offsets (2.480 vs.\ 2.095). Since their market shares are nearly identical, this suggests that A earns more profit per fill—likely by quoting at more favorable prices while still maintaining similar execution frequency. This indicates that Agent~A, having been explicitly trained under adversarial conditions, adopts bolder and more confident quoting strategies compared to B1, which is facing such an environment for the first time.

Overall, while Agent B1 was never explicitly trained under adversarial market conditions, its performance demonstrates a degree of adaptive capability—likely induced through repeated interaction with the robust Agent A during training. Nonetheless, Agent A consistently achieves superior outcomes, underscoring the limitations of indirectly acquired robustness. These results suggest that although strategic resilience can be partially transferred via structured multi-agent interaction, such transfer remains incomplete and does not substitute for direct adversarial exposure.

\subsection{Strategic Interaction and Market Effects in A–B1 Evaluation}
\begin{table*}[t]
\centering
\caption{Multi-Agent Performance: A vs B1, A vs B2, B1 vs B1, B1 vs B2 (mean ± std)}
\label{A_B1_B2}
\small
\renewcommand{\arraystretch}{1.1}
\resizebox{\linewidth}{!}{%
\begin{tabular}{lcccc}
\toprule
\textbf{Metric} & \textbf{A vs B1} & \textbf{A vs B2} & \textbf{B1 vs B1} & \textbf{B1 vs B2} \\
\midrule
PnL\_mean                 & 192.696 ± 19.703 / 274.555 ± 19.114 & 81.476 ± 11.535 / 260.953 ± 22.088 & 218.835 ± 19.353 / 222.811 ± 22.410 & 104.629 ± 11.040 / 233.721 ± 21.782 \\
Sharpe\_ratio            & 1.714 ± 0.127 / 1.772 ± 0.088 & 1.639 ± 0.113 / 1.769 ± 0.108 & 1.733 ± 0.104 / 1.748 ± 0.126 & 1.647 ± 0.097 / 1.754 ± 0.122 \\
Inventory\_volatility    & 19.068 ± 3.313 / 9.977 ± 2.944 & 6.290 ± 2.699 / 22.628 ± 3.158 & 14.494 ± 3.238 / 14.451 ± 3.025 & 4.473 ± 1.821 / 24.794 ± 2.837 \\
Quote\_aggressiveness    & 2.493 ± 0.011 / 2.126 ± 0.015 & 2.524 ± 0.028 / 1.482 ± 0.018 & 2.121 ± 0.011 / 2.119 ± 0.010 & 2.170 ± 0.030 / 1.476 ± 0.017 \\
Market\_share            & 0.373 ± 0.017 / 0.627 ± 0.016 & 0.162 ± 0.016 / 0.836 ± 0.016 & 0.497 ± 0.030 / 0.500 ± 0.030 & 0.243 ± 0.014 / 0.756 ± 0.014 \\
\midrule
Avg\_spread              & 4.619 ± 0.013 & 4.007 ± 0.025 & 4.239 ± 0.013 & 3.646 ± 0.034 \\
Price\_volatility        & 0.402 ± 0.155 & 0.402 ± 0.155 & 0.402 ± 0.155 & 0.402 ± 0.155 \\
Fill\_ratio              & 1.000 ± 0.001 & 0.999 ± 0.003 & 0.997 ± 0.002 & 0.999 ± 0.002 \\
Zero\_fill\_steps        & 27.000 ± 4.273 & 27.000 ± 4.273 & 27.000 ± 4.273 & 27.000 ± 4.273 \\
Joint\_drawdown\_ratio   & 0.104 ± 0.019 & 0.153 ± 0.034 & 0.121 ± 0.021 & 0.149 ± 0.024 \\
Herding\_ratio           & 0.000 ± 0.000 & 0.000 ± 0.000 & 0.467 ± 0.148 & 0.000 ± 0.000 \\
Inventory\_divergence    & 11.067 ± 4.768 & 17.130 ± 5.312 & 7.038 ± 2.759 & 21.383 ± 4.257 \\
Quote\_distance\_bid     & 0.189 ± 0.016 & 1.002 ± 0.052 & 0.012 ± 0.005 & 0.759 ± 0.022 \\
Quote\_distance\_ask     & 0.558 ± 0.029 & 1.082 ± 0.029 & 0.056 ± 0.017 & 0.630 ± 0.072 \\
Fill\_overlap\_ratio     & 0.208 ± 0.027 & 0.124 ± 0.024 & 0.219 ± 0.026 & 0.162 ± 0.023 \\
\bottomrule
\end{tabular}
}
\end{table*}

This section analyzes the joint evaluation of Agent A and Agent B1 in a fixed market environment (\(\lambda=400\), \(\sigma=1.1\)). The “A vs B1” column in Table~\ref{A_B1_B2} shows their respective performance in this setting, where each metric is presented as “A / B1”. Both agents are trained with self-interested objectives, with B1 trained via interaction with a pre-trained Agent A.

We examine whether strategic interactions between these two trained agents induce emergent behavioral patterns—such as behavioral convergence (e.g., quoting similarity), tacit coordination (e.g., implicit quote separation), or competitive divergence (e.g., inventory conflict or price overlap). Also, we assess how joint execution influences market outcomes—including liquidity, pricing efficiency (e.g., Avg\_spread), and agent profitability—relative to their single-agent baselines (“A\_Fix” and “B1\_Fix” in Table~\ref{tab:single_agent_eval}).

%\paragraph
\smallskip \emph{Profitability and Execution Dominance.}
According to Table~\ref{A_B1_B2}, Agent~B1 achieves a higher mean PnL (274.555 vs.\ 192.696) and slightly better Sharpe ratio (1.772 vs.\ 1.714) than Agent~A, despite A outperforming B1 in both metrics when tested alone (PnL = 506.49 vs.\ 437.293). Their combined PnL in the joint setting (467.250) exceeds B1’s solo result but trails A’s, suggesting that B1 not only dominates A in this setting but also extracts more profit per fill than in isolation. B1 also captures significantly more execution share (0.627 vs.\ 0.373), implying that Agent A is partially crowded out. These results suggest that B1 assumes a dominant execution role in the joint environment and extracts greater profitability.

%\paragraph
\smallskip \emph{Agent Behavior and Market Impact.}
In the joint environment, both Agent~A and Agent~B1 show reduced inventory volatility compared to their single-agent baselines (28.335), indicating more conservative inventory management. The decline is more pronounced for B1 (down to 9.977 vs. A’s 19.068), suggesting stronger risk control and avoidance of extreme inventory swings. Despite this, their inventory divergence remains substantial (11.067), and the fill overlap is low (0.208), pointing to distinct execution paths. A herding ratio of 0.000 and quote distances of 0.189 (bid) and 0.558 (ask) further indicate divergent quoting strategies, especially on the ask side.

Overall, although both agents are self-interested profit maximizers, there is no evidence of collusion. Agent~B1 dominates in profitability and execution share, while Agent~A exhibits distinct quoting and inventory behavior, suggesting that each agent competes independently. This contrasts with findings in retail pricing settings~\cite{calvano2020artificial}, where artificial agents have been shown to engage in collusion—likely because the simulated market price here is driven by an exogenous stochastic process, preventing price manipulation by the agents.

\subsection{Strategic Interaction and Market Effects in A–B2 Evaluation}
This section analyzes the joint evaluation of Agent~A and Agent~B2 in a fixed market (\(\lambda = 400\), \(\sigma = 1.1\)). Table~\ref{A_B1_B2} (“A vs B2” column) reports their performance, with metrics shown as “A / B2”. Agent~A is adversarially trained for robustness and generalization, while B2 is explicitly optimized to suppress A’s profitability in joint execution. Unlike general-purpose agents, B2 is not profit-driven, but trained to exert persistent competitive pressure. This evaluation examines (1) how a robust agent like A performs under sustained adversarial pressure, and (2) how their interaction shapes execution dynamics and market outcomes.

%\paragraph
\smallskip \emph{Profitability and Execution Dominance.}
As shown in Table~\ref{A_B1_B2}, Agent~A achieves a much lower mean PnL than B2 (81.476 vs.\ 260.953), in sharp contrast to the “A vs.~B1” setting, where A outperformed with a PnL of 192.696. The disparity extends to execution share: A secures only 16.2\% of total executions, while B2 captures 83.6\%, reflecting a consistent and substantial suppression of A’s market presence. A’s Sharpe ratio also drops to 1.639—below both its solo benchmark (1.749 in \texttt{A\_Fix}) and B2’s value in the joint setting (1.769)—indicating that B2 imposes a more severe impact on A’s profitability than B1.

Agent~A also quotes significantly wider spreads (2.524 vs.\ B2’s 1.482), even wider than its solo setting (2.480), suggesting a more conservative pricing strategy under pressure. In contrast, B2 places tighter quotes near the mid-price, offering more competitive prices. By narrowing its spread, B2 sacrifices per-trade margin for higher fill probability—effectively “paying for priority” to capture order flow. This aggressive execution allows B2 to crowd out A and dominate profitability in a zero-sum environment.

%\paragraph
\smallskip \emph{Agent Behavior and Market Impact.}
The herding ratio remains zero, and quote distances on both bid and ask sides are higher than in other evaluated pairings, indicating that Agent~A and B2 follow distinctly different quoting strategies. Inventory divergence (17.130) further reflects divergent risk-taking behavior: B2 holds larger and more volatile inventory positions (inventory volatility = 22.628), while A remains conservative (6.290), consistent with its wider quotes. The low fill overlap ratio (0.124) confirms that the agents rarely compete for the same trades and largely operate in disjoint execution zones.

The joint drawdown ratio increases to 0.153 (vs.\ 0.104 in A vs.~B1), indicating more frequent simultaneous losses. This suggests the emergence of a mutually unprofitable dynamic—likely driven by structural conflict rather than coordination. Supporting this, the combined PnL of A and B2 (342.429) is substantially lower than A’s solo performance (506.490), implying that adversarial interaction fails to generate additional market value.

Nonetheless, the A–B2 pairing results in a narrower average spread (4.007 vs.\ 4.959 in \texttt{A\_Fix}), suggesting improved market liquidity. However, this systemic gain comes at the expense of A’s individual performance, highlighting a fundamental trade-off between market-level efficiency and agent-level robustness under competitive pressure.

Overall, the A–B2 evaluation reveals a one-sided competitive dynamic: Agent~B2 dominates execution and profitability through aggressive pricing, while Agent~A retreats into conservative behavior with limited market share. Despite the lack of coordination, their interaction induces mutual inefficiencies—reflected in elevated drawdown and reduced total PnL—suggesting that persistent adversarial pressure can degrade overall value creation. Although market liquidity improves, it does so at the cost of A’s robustness, illustrating the tension between individual resilience and systemic efficiency in competitive multi-agent settings.

\subsection{Strategic Interaction and Market Effects in B1–B2 Evaluation}

This section analyzes the strategic interaction between Agent~B1 and Agent~B2 in a fixed market (\(\lambda = 400\), \(\sigma = 1.1\)). Agent~B1 is a self-interested agent trained through interaction with Agent~A to maximize its own profitability. In contrast, B2 is an adversarial agent—originally trained to suppress A, but reconfigured here to target B1’s PnL instead.

The “B1 vs B2” column in Table~\ref{A_B1_B2} reports their performance, with metrics shown as “B1 / B2”. This setup enables us to assess how a self-optimizing agent responds to targeted suppression and how such interaction shapes execution dynamics and market outcomes.

%\paragraph
\smallskip \emph{Profitability and Execution Dominance.}
As shown in Table~\ref{A_B1_B2}, Agent~B2 achieves substantially higher profitability than B1 (233.721 vs.\ 104.629), indicating dominance in the joint environment. This advantage extends to execution share: B2 captures 75.6\% of total fills, while B1 secures only 24.3\%, showing that B2 consistently outcompetes B1 for order flow. Sharpe ratios reinforce this gap, with B2 achieving 1.754 compared to B1’s 1.647—highlighting superior risk-adjusted performance.

In terms of quoting strategy, B2 posts tighter quotes (aggressiveness = 1.476) than B1 (2.170), while B1’s aggressiveness is slightly higher than in its solo setting (2.095 in \texttt{B1\_Fix}). The overall market spread narrows in this pairing (3.646 vs.\ 4.189 in \texttt{B1\_Fix}), suggesting that B2's tighter offset contributes to improved market liquidity. This aggressive pricing enables B2 to quote near the mid-price, boosting its fill probability and execution dominance. In contrast, B1’s wider spreads and reactive posture lead to reduced market participation and lower profitability.

%\paragraph
\smallskip \emph{Agent Behavior and Market Impact.}
The herding ratio remains zero, and quote distances on both bid (0.759) and ask (0.630) sides are relatively high, indicating a lack of quoting alignment between the two agents.

Inventory behavior further highlights strategic divergence. B1’s inventory volatility drops sharply from 28.335 in solo evaluation to just 4.473, while B2 maintains a high level (24.794). Inventory divergence between the two agents is substantial (21.383), reflecting B2’s aggressive, flow-seeking strategy and B1’s conservative positioning. The fill overlap ratio is moderate (0.162)—lower than in A--B1 (0.208) but higher than in A--B2 (0.124)—suggesting that B1 and B2 occasionally compete for fills, but generally operate in partially disjoint execution zones.

Compared to its interaction with Agent~A, B1 exhibits less overlap in liquidity targeting when facing B2. This likely reflects their divergent quoting behaviors: B2 dominates near the mid-price, while B1 quotes more defensively at wider spreads. The quote distances across all three pairings support this interpretation.

Overall, the B1–B2 evaluation reveals a clear competitive asymmetry: B2 dominates both execution and profitability, while B1, despite being self-interested and previously robust against A, fails to assert itself against a strategically focused and aggressive opponent.

We attribute B2’s dominance over B1 to two possible factors. First, B2’s reward shaping promotes inherently suppressive, low-entropy strategies aimed at neutralizing opponents—regardless of identity. Second, B2 acts as a game-shaper: while B1 optimizes for market-based equilibrium under self-interest, B2 tends to prioritize execution control, which may come at the expense of cooperative dynamics or market balance. This structural advantage becomes particularly evident when comparing B1's profitability across matchups. As shown in the ``B1 vs B1'' column of Table~\ref{A_B1_B2}, when two agents with highly similar quoting behavior interact, both achieve relatively high individual PnLs (218.835 and 222.811), yielding a combined total of 441.65. In contrast, in the B1--B2 pairing, although the total PnL drops sharply to 338.35, this decline is entirely driven by B1’s profit falling to 104.629, while B2’s rises to 233.721. Notably, B2 not only suppresses its opponent but also outperforms both B1 agents from the symmetric setting. This contrast underscores that B2’s dominance does not stem from exploiting a weaker counterpart, but from the inherent competitiveness and control-oriented nature of its policy.

\subsection{Strategic Interaction and Market Effects in A–B$^\star$ and B1–B$^\star$ Evaluation}

This section analyzes the interaction between baseline agents (A, B1) and the hybrid market-making agent B$^\star$ in a fixed market (\(\lambda = 400\), \(\sigma = 1.1\)). Agents~A and B1 are trained under self-interested objectives to maximize individual profitability. In contrast, B$^\star$ is trained through interaction with Agent~A, learning to balance self-interested maximization with suppressing its counterpart’s performance. Its objective explicitly penalizes A’s profitability, instilling adversarial tendencies into its policy.

In the evaluations presented here, B$^\star$ is paired with both A and B1, the latter being previously unseen during training. This setup is designed to assess whether B$^\star$ exhibits generalized suppressive behavior or adaptively modulates its strategy based on the nature of its opponent.

\begin{table}[t]
\centering
\caption{Multi-Agent Performance: A vs $\text{B}^\star$ and B1 vs $\text{B}^\star$ (mean ± std)}
\label{tab:bstar_summary}
\scriptsize
\renewcommand{\arraystretch}{1.1}
\resizebox{\linewidth}{!}{%
\begin{tabular}{lcc}
\toprule
\textbf{Metric} & \textbf{A vs $\text{B}^\star$} & \textbf{B1 vs $\text{B}^\star$} \\
\midrule
PnL\_mean & 172.469 ± 16.464 / 285.461 ± 26.194 & 215.023 ± 19.933 / 218.830 ± 24.131 \\
Sharpe\_ratio & 1.775 ± 0.123 / 1.742 ± 0.103 & 1.860 ± 0.117 / 1.647 ± 0.122 \\
Inventory\_volatility & 10.116 ± 3.555 / 18.983 ± 3.642 & 6.831 ± 2.015 / 22.244 ± 2.080 \\
Quote\_aggressiveness & 2.534 ± 0.032 / 2.100 ± 0.020 & 2.152 ± 0.017 / 2.127 ± 0.021 \\
Market\_share & 0.329 ± 0.021 / 0.670 ± 0.021 & 0.471 ± 0.028 / 0.526 ± 0.028 \\
Mean\_omega ($\text{B}^\star$) & - / 0.697 ± 0.002 &  - / 0.695 ± 0.002  \\
\midrule
Avg\_spread & 4.634 ± 0.031 & 4.280 ± 0.025 \\
Price\_volatility & 0.402 ± 0.155 & 0.402 ± 0.155 \\
Fill\_ratio & 0.999 ± 0.001 & 0.997 ± 0.003 \\
Zero\_fill\_steps & 27.000 ± 4.273 & 27.000 ± 4.273 \\
Joint\_drawdown\_ratio & 0.107 ± 0.026 & 0.115 ± 0.023 \\
Herding\_ratio  & 0.000 ± 0.000 & 0.000 ± 0.000  \\
Inventory\_divergence & 11.293 ± 5.570 & 16.373 ± 3.319 \\
Quote\_distance\_bid & 0.463 ± 0.049 & 0.252 ± 0.010 \\
Quote\_distance\_ask & 0.477 ± 0.025 & 0.181 ± 0.033 \\
Fill\_overlap\_ratio & 0.192 ± 0.025 & 0.208 ± 0.027 \\
\bottomrule
\end{tabular}
}
\end{table}

Table~\ref{tab:bstar_summary} reports performance metrics for both A–B$^\star$ and B1–B$^\star$ pairings, with entries formatted as “A / B$^\star$” and “B1 / B$^\star$”, respectively. This dual evaluation addresses three core questions:

\begin{enumerate}[nosep, left=0pt]
    \item Whether the hybrid agent B$^\star$ exhibits consistent behavioral tendencies across distinct market contexts—that is, does it lean toward self-interested behavior, competitive behavior, or display adaptive modulation between the two?
    \item How does B$^\star$ interact with distinct self-interested agents (A vs.~B1)—through suppression, implicit cooperation, or role differentiation?
    \item Compared to single-agent baselines (A\_Fix, B1\_Fix in Table~\ref{tab:single_agent_eval}), how does B$^\star$ affect market structure and agent-level outcomes?
\end{enumerate}

%\paragraph
\smallskip \smallskip \emph{Profitability and Execution Dominance.}
B$^\star$ maintains tighter quoting across both pairings (quote aggressiveness: 2.100 vs. 2.534 against A; 2.127 vs. 2.152 against B1), allowing it to position closer to the mid-price and enhance execution priority. This assertive quoting behavior is accompanied by significantly higher inventory volatility (18.983 against A; 22.244 against B1), indicating a more active and risk-tolerant trading posture. Quote distance metrics further contextualize this interaction. In the B1 pairing, the average market quote distance (bid/ask = 0.252 / 0.181) is narrower than in the A pairing (0.463 / 0.477). Since B$^\star$ is present in both scenarios, this difference likely reflects B1’s tighter quoting behavior compared to A, rather than a shift induced by B$^\star$. This also suggests that B$^\star$ and B1 exhibit greater similarity in quoting strategy than B$^\star$ and A—echoing earlier observations that A and B1 are more similar to each other than B1 and B2, while B1 and B2 are still more aligned than A and B2. These patterns may stem from the three-tier agent architecture employed in this work. Consequently, B$^\star$ is operating in a more competitive liquidity environment when facing B1, which may partly explain its heightened inventory volatility in that setting.

Such behavior likely contributes to B$^\star$'s ability to dominate execution: it captures 67.0\% of total market share against A and 52.6\% against B1, reflecting consistent control over order flow. The closer balance in the B1 pairing suggests a more symmetric interaction, whereas the skew toward B$^\star$ in the A pairing may reflect either A’s limited responsiveness or a strategic asymmetry in favor of B$^\star$. In terms of profitability, B$^\star$ outperforms both A and B1 in mean PnL (285.461 vs. 172.469 against A; 218.830 vs. 215.023 against B1), confirming its execution edge translates into economic gain. However, its risk-adjusted performance is mixed: against A, its Sharpe ratio (1.742) is close to A’s (1.775), while against B1, it drops to 1.647—noticeably below B1’s 1.860—suggesting greater return volatility despite nominal profit advantage.

%\paragraph
\smallskip \emph{Agent Behavior and Adaptive Strategy.}
Behavioral indicators suggest that B$^\star$ interacts competitively while maintaining independence across both matchups. The fill overlap ratios (0.192 with A, 0.208 with B1) are notably higher than those observed in B2 pairings, indicating more frequent contention for the same trades. This suggests that B$^\star$ does not avoid its counterpart's liquidity zones, but rather engages in active trade-level interaction. Despite this overlap, B$^\star$ avoids mimicry. Inventory divergence remains moderate (11.293 vs. A; 16.373 vs. B1), indicating differentiated risk-taking behavior even under shared execution pressure. Supporting this, the herding ratio is zero, confirming the absence of quote matching or strategic alignment.

These results suggest that B$^\star$ adapts to its counterpart’s behavior while maintaining distinct execution patterns. It competes actively for flow without converging structurally, balancing responsiveness with strategic independence.

%\paragraph

\smallskip \emph{Market-Level Effects.}
Relative to single-agent benchmarks (A\_Fix and B1\_Fix in Table~\ref{tab:single_agent_eval}), the introduction of B$^\star$ produces mixed effects on both market structure and agent-level outcomes. In A–B$^\star$, the average spread narrows from 4.959 to 4.634, suggesting improved liquidity; however, in B1–B$^\star$, it widens slightly from 4.190 to 4.280, indicating a marginal decline in efficiency. This inconsistency may reflect the fact that B$^\star$ is not trained toward a fixed market-making goal but instead learns to balance between maximizing its own returns and suppressing those of its counterpart. As a result, its strategic behavior introduces variability into the market environment, producing different structural outcomes depending on the opponent. Meanwhile, the combined PnL in each pairing declines relative to the solo-agent baselines: A and B$^\star$ jointly earn 457.93, compared to 506.490 from A alone; B1 and B$^\star$ total 433.853, slightly below B1’s solo performance of 437.293. This indicates that B$^\star$’s presence does not unlock additional market value—rather, it compresses total profitability and imposes systemic pressure on both the counterpart and the overall trading environment.

Overall, B$^\star$ leverages relatively small quoting advantages to consistently capture greater market share, gradually eroding the profitability of both A and B1. However, its presence does not unlock additional market value; instead, total returns decline relative to single-agent baselines. This suggests that while B$^\star$ competes effectively at the agent level, it introduces varying degrees of pressure on counterpart strategies and overall market efficiency.

\subsection{Comparative Summary: B2 vs. B$^\star$}

Both B2 and B$^\star$ demonstrate strong competitive capabilities in multi-agent markets, but their behavioral structures and underlying incentive mechanisms differ fundamentally.

\textbf{B2 is a zero-sum suppression-oriented agent}, whose reward is explicitly defined as the negative PnL of its opponent ($-\text{PnL}_\text{opponent}$). Trained specifically to suppress Agent A, B2 evolves into a consistently adversarial policy. This results in severe performance degradation for its counterparts (e.g., A’s mean PnL drops to 81.476, and B1’s to 104.629), while B2 itself maintains dominant execution statistics—over 75\% market share and highly aggressive quoting behavior (e.g., average spread in B1–B2 narrows to 3.646). However, such behavior may come at the cost of increased conflict and systemic tension.

\textbf{B$^\star$, by contrast, is an incentive-adaptive agent} trained to balance between self-interested and competitive via a tunable weight $\omega$. When interacting with non-adversarial agents like A and B1, B$^\star$ tends to adopt a more self-interested stance (as indicated by $\omega \approx 0.696$), focusing on maximizing its own return without aggressively targeting counterpart losses. Despite this, B$^\star$ consistently outperforms its counterparts in both PnL and market share while allowing A and B1 to retain relatively stable Sharpe ratios and risk profiles. Compared to B2, B$^\star$ achieves market dominance through less aggressive means, promoting a healthier interaction dynamic.

\textbf{In summary}: The competitive agent (B2) achieves gains through rigid adversarial behavior but induces stress in market dynamics; the hybrid agent (B$^\star$) leverages flexible modulation between self-interested and competitive objectives to attain modest superiority while preserving systemic stability. In heterogeneous agent markets, the hybrid agent’s behavioral adaptability suggests a more sustainable and strategically viable path for long-term coexistence.

\section{Conclusion}

This paper presents a hierarchical multi-agent reinforcement learning framework designed to study the structural interplay between heterogeneous trading agents in competitive markets. By introducing explicitly configured agent roles—self-interested, competitive, and incentive-modulating—we enable controlled evaluation of behavioral asymmetry and adaptive response. A suite of interaction-level metrics is proposed to support interpretable analysis of emergent structure and strategy impact beyond aggregate performance.

Our comparative analysis reveals that rigid suppression strategies, as implemented by Agent~B2, can outperform self-interested agents in zero-sum scenarios by monopolizing order flow and tightening execution spreads. However, such dominance often disrupts competitive balance. In contrast, the hybrid agent B$^\star$—despite exhibiting self-interested tendencies when interacting with other profit-seeking agents—demonstrates strategic adaptability by securing market share with less severe impact on competitors. This contrast highlights the potential of behavioral modulation in enabling sustainable coexistence among learning agents.

While our framework does not induce emergent coordination or imitation, it provides a structured basis for identifying interaction-level patterns across heterogeneous agents. One aspect of the current formulation is that the hybrid agent’s modulation parameter \( \omega \) is directly optimized as part of the action vector and remains tightly coupled to the reward during training. As a future direction, one could pursue a meta-learning formulation where agents learn to adapt their objective preferences based on environmental conditions or opponent behavior. Such meta-level control could support more generalizable incentive adaptation across diverse market regimes.

\bibliographystyle{ACM-Reference-Format}

\bibliographystyle{unsrt}
\bibliography{references}
\end{document}